# Manipulating Spin Chirality of Magnetic Skyrmion Bubbles by In-Plane Reversed Magnetic Fields in $(Mn_{1-x}Ni_x)_{65}Ga_{35}$ (x = 0.45) magnet

Bei Ding[1,3], Jie Cui[1,3], Guizhou Xu[2], Zhipeng Hou[1], Hang Li[1,3], Enke Liu[1], Guangheng Wu[1], Yuan Yao[1*], and Wenhong Wang[1,4*]

[1] Beijing National Laboratory for Condensed Matter Physics and Institute of Physics, Chinese Academy of Sciences, Beijing 100190, China

[2] School of Materials Science and Engineering, Nanjing University of Science and Technology, Nanjing 210094, China

[3] University of Chinese Academy of Sciences, Beijing 100049, China

[4] Songshan Lake Materials Laboratory, Dongguan, Guangdong 523808, China

To whom correspondence should be addressed E-mail: yaoyuan@iphy.ac.cn or wenhong.wang@iphy.ac.cn

**ABSTRACT:** Understanding the dynamics of the magnetic skyrmion, a particle-like topologically stable spin texture, and its response dynamics to external fields are indispensable for the applications in spintronic devices. In this letter, the Lorentz transmission electron microscopy (LTEM) was used to investigate the spin chirality of the magnetic skyrmion bubbles (SKBs) in the centrosymmetric magnet MnNiGa at room temperature. The reversal of SKBs excited by the in-plane magnetic field has been revealed. Moreover, the collective behavior of interacting spin chirality can be manipulated by reversing the directions of the magnetic fields on a wedge-shaped thin plate. The dynamic behavior of the bubbles at different position of the thin plate has been explored with the micromagnetic simulation, indicating a non-uniform and nontrivial dynamic magnetization on the surfaces and center of the thin plate during the spin chirality reversal. The results suggest that the controllable symmetry breaking of the SKBs arising from thickness variation provides an ability to manipulate the collective behavior of the spin chirality with small external fields, leading to a promising application in nonvolatile spintronic devices for magnetic skyrmions.

# Ⅰ. INTRODUCTION

Skyrmions, the particle-like topologically stable spin textures, are usually stabilized by the Dzyaloshinskii-Moriya interaction (DMI) in non-centrosymmetric magnets with broken inversion symmetry, such as MnSi [1-3], FeGe [4,5], FeCoSi [6], $Cu_2OSeO_3$ [7-9], and Mn-Pt-Sn [10]. The response dynamic of magnetic skyrmions to external stimuli is an important property for the manipulation of skyrmions, which is indispensable for practical applications in spintronic. The stability of skyrmions has been investigated by using various external field, including electromagnetic wave resonance [11], cooling [12], anisotropic mechanical strain [13] and electric field [14,15]. Moreover, the control of skyrmions chirality has been demonstrated by means of the chemical composition [16], grain orientation [5,17], and in-plane magnetic field [18]. Similar behaviors are expected for other topologically nontrivial spin textures known as magnetic skyrmion bubbles (SKBs) in some centrosymmetric magnets with dipole-dipole interaction and uniaxial magnetic anisotropy, such as MnNiGa [19], $Ni_2MnGa$ [20], $Fe_3Sn_2$ [21] and Ba $(Fe_{1-x-0.05}Sc_xMg_{0.05})_{12}O_{19.16}$ [22]. Contrary to DMI-stabilized skyrmions with a fixed chirality, SKBs in centrosymmetric magnets possess two degrees of freedom, i.e., vorticity and helicity [23], making them suitable to manipulate spin chirality and topological texture through stimuli.

Recent simulation demonstrates that the chirality reversal of SKBs could be controlled by tuning driving current density [24] which exhibits potential applications as a binary memory device. In addition, thermal activation can also reverse the SKBs helicity [25], although it is not suitable for practical applications. Yu *et al* [22,26] found that the chirality reversal occurred while tilting the applied magnetic field but the demand for out-of-plane bias or low temperature limits its realization in the future devices. Furthermore, the field free biskyrmion bubbles have been explored in centrosymmetric hexagonal MnNiGa alloys at room temperature via appropriate field cooling (FC) procedure [19,27]. In this work, motivated by the above-mentioned approaches, the manipulation of the spin chirality of the field free skyrmion bubble lattice (SKBL) generated within wedge-shaped $(Mn_{1-x}Ni_x)_{65}Ga_{35}$ (x = 0.45) thin plate was demonstrated by in situ LTEM with the help of in-plane magnetic field at room temperature. Micromagnetic simulation reproduced the collective reversal behavior of the SKB chirality. Moreover, the dynamic transition at different position of the thin plate has been investigated, indicating a non-uniform and nontrivial dynamic magnetization on the surfaces and center of the thin plate during the spin chirality reversal.

## Ⅱ. METHODS

### A. Sample preparation

Experiments were performed on a polycrystalline $(Mn_{1-x}Ni_x)_{65}Ga_{35}$ (x = 0.45) sample, synthesized by arc-melting mixtures of high purity Mn, Ni and Ga in a pure argon atmosphere. The crystal structure of the alloy was studied using powder X-ray diffraction measurements, which demonstrated a hexagonal structure with a space group of $P6_3/mmc$ (No. 194) (Fig. 1(a)). The Curie temperature $T_c$ is determined to be 345 K from the temperature dependence of magnetization. The singular point detection (SPD) technique [28-30] was applied to calculate the anisotropy of the oriented polycrystalline $(Mn_{1-x}Ni_x)_{65}Ga_{35}$ (x = 0.45) samples. The hysteresis loops were measured under a series of the field (50 Oe, 100 Oe, 150 Oe and 200 Oe) with the bulk sample reflecting a weak coercivity (see S1 within the Supplemental Material [31]).

### B. Sample thickness measurement

The thin plate for the LTEM observation was cut from the polycrystalline bulk and thinned by the mechanical polishing and argon ion milling, which represents a wedge-like variation in the thickness (see S2 within the Supplemental Material [31]). The thickness of the sample region was measured by using electron energy-loss spectroscopy (EELS) log-ratio techniques with a Titan G2 60–300 (FEI). The normalized thickness ($t/\lambda$) (t represents thickness, $\lambda$ means the free path for inelastic scattering of the electron beam in the material) map is together with a series of EELS spectra extracted from the spectrum image. By taking account of the spectrum collection conditions ($E_0$ = 300 *KV*, $\alpha$ = 24 *mrad*, $\beta$ = 8.6 *mrad*), the inelastic mean free paths $\lambda$ of sample was calculated.

### C. LTEM measurements

The magnetic domain contrast was made by using Titan G2 60-300 in the Lorentz

TEM mode equipped with a double-tilt heating holder and FEI JEOL-dedicated Lorentz TEM equipped with a magnetic field holder. To obtain zero free SKBL, a double-tilt heating holder (Model 652, Gatan Inc.) with a smart set hot stage controller (Model 901, Gatan Inc.) was used to raise the specimen temperature from 300 K to 400 K. The specific FC manipulation was shown as follows: first, the sample was heated up to 360 K, which is higher than Curie temperature $T_C \sim 345$ K. Second, a small perpendicular magnetic field of 500 Oe was applied by increasing the objective lens current gradually in very small increments. Third, the temperature of sample was cooled down gradually from 360 K to 300 K. Finally, at 300 K, the small perpendicular magnetic field was turned off. The experimental process was recorded by the charge coupled device (CCD) camera (see Movie 1 within the Supplemental Material [31]). A FC procedure here is just needed for the thin plate to generate the SKBs but does not affect the subsequent manipulation of spin chirality of magnetic SKBs by reversing in plane magnetic fields (see S3 within the Supplemental Material [31]).

The in-plane magnetic field applied to switch the chirality of the SKBL was induced by the excitation coil in the magnetic field holder. To determine the chirality of SKBs, three sets of the images (under-, over- and in-focus) were acquired by using a CCD camera and the in-plane magnetization distribution was obtained by using the Qpt software based on the transport-of-intensity equation (TIE) [32].

**D. Micromagnetic simulations**

The micromagnetic simulations were carried out by using 3D object oriented Micromagnetic framework (OOMMF) code [33]. The magnetization obtained from OOMMF simulation was input into a homemade Digitalmicrograph (DM) script to simulate the LTEM images [34] (see S4 within the Supplemental Material [31]). The material parameters were obtained from the experiments on $(Mn_{1-x}Ni_x)_{65}Ga_{35}$ (x = 0.45) as follows: the saturation magnetization $M_S = 8\times10^5$ A/m$^3$, the magneto-crystalline anisotropy constant $K_u = 2.65\times10^5$ J/m$^3$, which was measured by SPD technique. The exchange constant $A = 1.4\times10^{-11}$ J/m by $D = \pi\sqrt{A/K_u}$, where D is the domain wall width obtained from the LTEM results.

First, for the analysis of skyrmion bubble chirality reversal dynamics, only a single skyrmion bubble is defined with a cylindrical geometry with experimentally observed structures. Slab geometries of dimensions 192 nm x 192 nm x (120~130 nm) were discretized with 1.5 nm tetrahedra. The magnetic field was applied along the x direction to transform the SKB into achiral bubble. Then, SKBL was simulated by using the slab with a volume of 1 μm × 1 μm ×100 nm and a wedge thickness varied from 100 nm to 130 nm with 6 nm x 6 nm x 2 nm mesh cell size. The quantity of C = +1 and C = -1 SKBs dependences of the applied in-plane magnetic field times was investigated when the in-plane along x or y direction of the SKBL.

## Ⅲ. RESULT AND DISSCUSSION

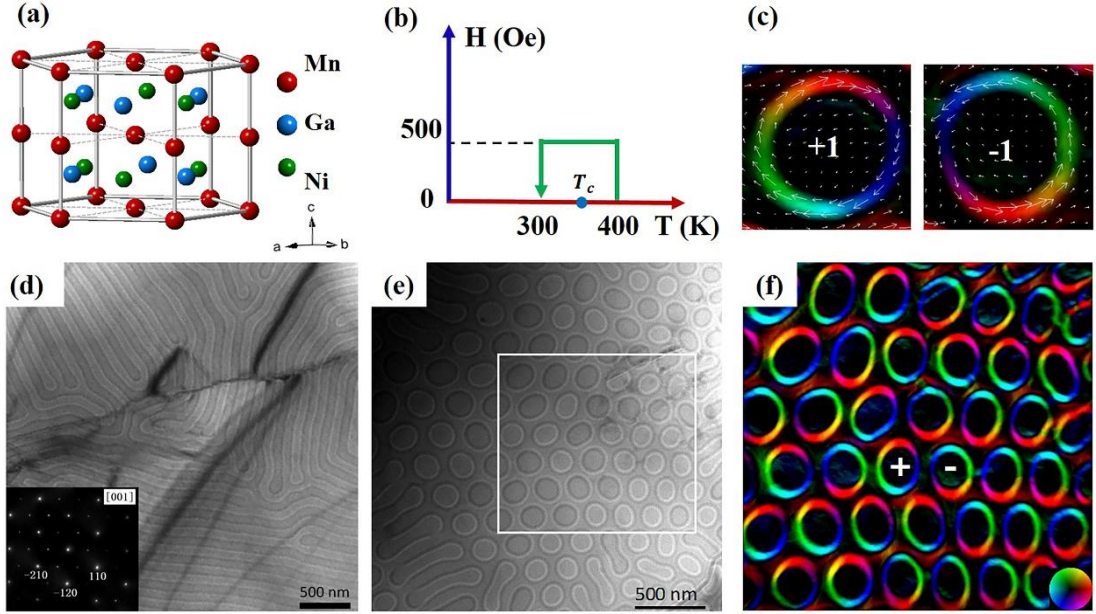

**FIG. 1.** Structure and magnetic domain textures observed in the $(Mn_{1-x}Ni_x)_{65}Ga_{35}$ thin plate with x=0.45. (a) Schematic crystal structures of a layered $Ni_2In$-type structure Mn-Ni-Ga with alternating stacks of Mn atomic layers and Ni-Ga atomic layers along the c-axis. (b) Schematic illustration of the field cooling (FC) procedure. (d) Magnetic stripe domain textures observed at room temperature along [001] axis. The inset shows the selected area electron diffraction pattern (SEAD) for a single-crystalline domain. (e) Real-space observation of zero magnetic field room temperature skyrmion bubbles (C = ±1) and magnetic stripe after FC manipulation. (f) In-plane magnetization of the selected skyrmion bubbles from (e). (c) Magnified images of two skyrmion bubbles with opposite chirality indicated in (f). Color wheel and arrows indicate the magnitude and direction of the in-plane magnetization. "+" and "-" represent the clockwise (C = 1) and counterclockwise (C = -1) chirality of the skyrmion bubbles.

Various magnetic domains were observed on the wedge-shaped alloy specimen [35], in the present study, we only focus on the magnetic stripes along the [001] axis as shown in Fig. 1(d). Based on the previous work, a FC procedure (see Fig. 1(b)) was applied on the thin plate to generate the SKBL [27]. Figure 1(e) represents an over-focus LTEM image of SKBL at room temperature and zero magnetic field, coexistent with the stripe domains. To characterize the in-plane topological magnetization texture in detail, we have analyzed the selected SKBL LTEM images by using TIE technique (Fig. 1(f)). The magnified topological textures in Fig. 1(c) clearly indicate that two SKBs with the opposite chirality. Accordingly, the dark/white SKBs shown in Fig. 1(e) can be denoted as the clockwise "+" and anticlockwise "-", respectively. In addition, the achiral bubbles are observed in the subsequent work (see Fig. 2) and corresponding magnetizations are shown in Fig. S4 (see S4 within the Supplemental Material [31]). A chirality C is used to clarify these different bubbles: chiral bubbles with the in-plane magnetization rotate either clockwise (C = +1) or counterclockwise (C = -1), while the topological achiral bubbles with C = 0.

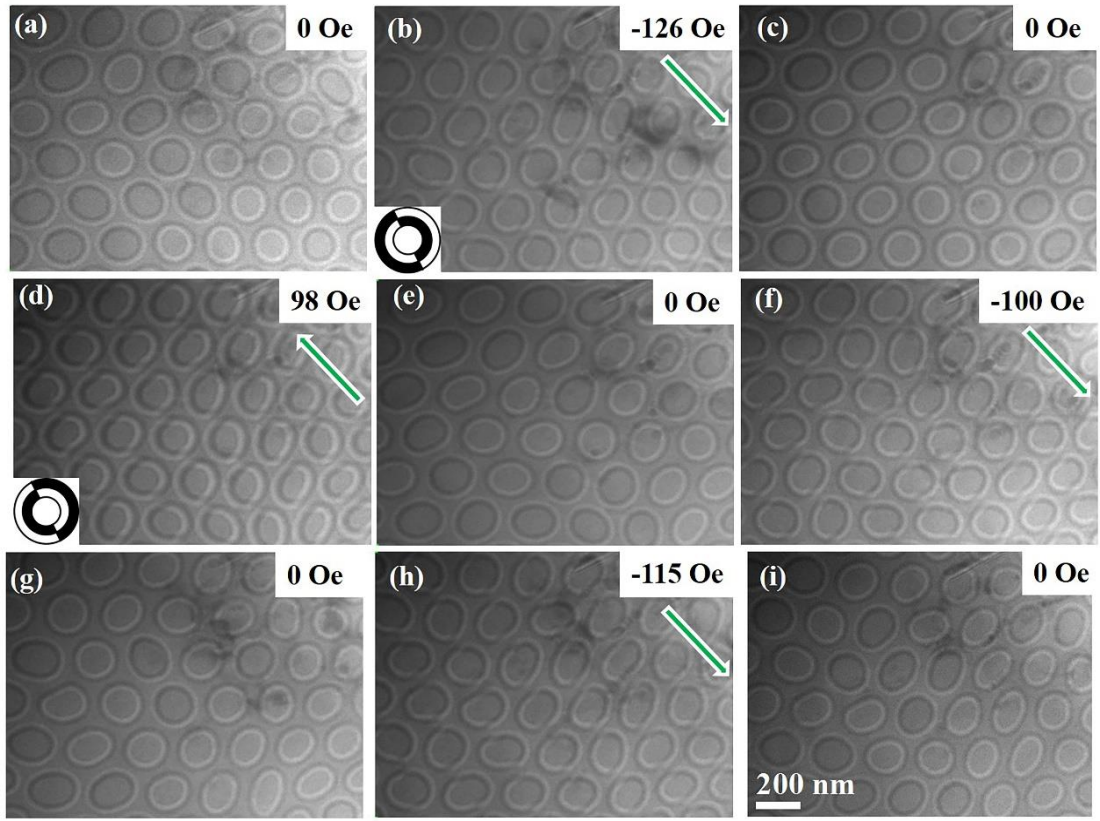

**FIG. 2.** Real-space observation of chirality transformation between skyrmion bubbles (C= ±1) and achiral bubbles (C = 0) induced by the applied in-plane magnetic field. (a) - (i) Bubbles observed at room temperature with various in-plane magnetic field. The inset in (b) and (d) show two kinds of achiral bubbles with opposite in-plane magnetic field. The green arrow with white rim indicates the direction of the in-plane magnetic field.

Figures 2(a)-2(i) illuminate the chirality switch of SKBs in a magnetic field parallel to the thin plate surface. The direction of the in-plane magnetic field is marked by the green arrow with white rim. With the enhancement of the magnetic field, the SKBs with C = ±1 were converted into the achiral bubbles (C = 0), as shown in Fig. 2(b). When the magnetic field decreased to zero, the achiral bubbles was transformed back into the chiral SKBs (Fig. 2(c)). This can be attributed to the fact that the energy of the SKBs is lower compared with the achiral bubbles [24]. As the field direction was reversed, the achiral bubbles appeared again (Fig. 1(d)). Once fading of the field to zero led to the recovery of the chiral SKBs (Fig. 1(e)). This process could be circled forth and back many times (Figs. 2(a-i)).

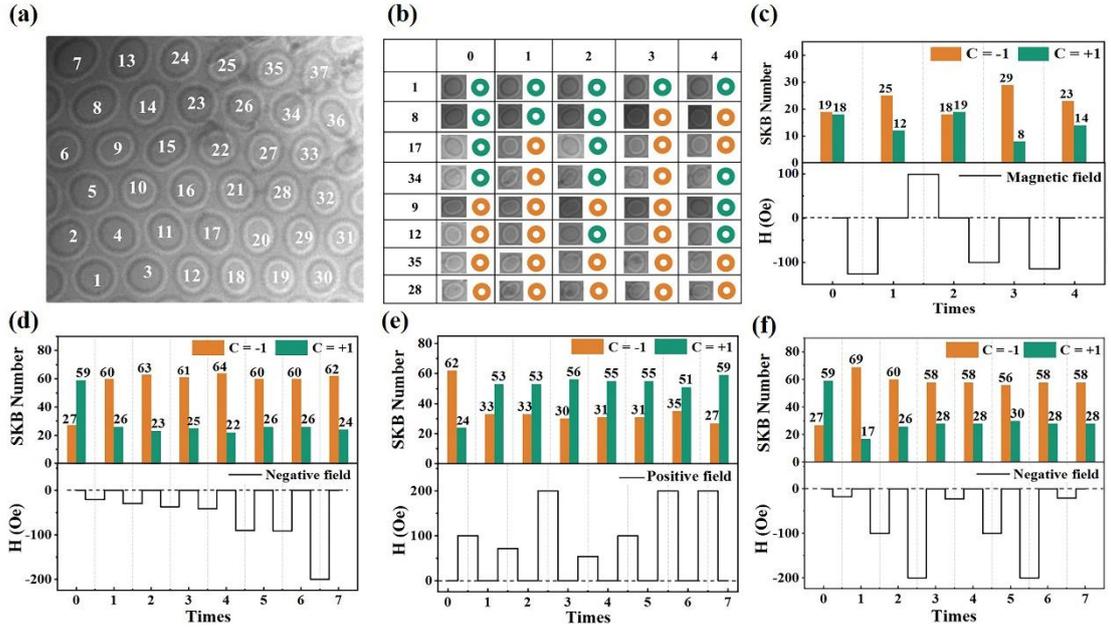

**FIG. 3**. Statistics of chirality reversal obtained after applied in-plane field. (a) LTEM image was obtained at room temperature under zero magnetic field which was considered as the initial state. (b) The chirality reversal of some SKBs at different field conditions. The numbers in first column are the labels of bubbles in (a) and the numbers in first row represent the corresponding magnetic field condition in (c). The green and orange circles indicate the clockwise (C = +1) and counter-clockwise (C = -1) chirality of the skyrmion bubbles, respectively. (c) The upper histogram shows the statistics number of C = ±1 skyrmion bubbles after application of the in-plane magnetic field. The orange and green bands indicate the numbers of the C = ±1 skyrmion bubbles, respectively. The corresponding number represents the value of C = +1 and C = -1 skyrmion bubbles. The low chart is the applied magnetic field. (d)- (e) The statistics of C = ±1 skyrmion bubbles after only applied positive or negative magnetic field, respectively.

With the intention of clearly analysis the quantity of reversed SKBs, 37 SKBs in Fig. 3(a) were tracked. Figure 3(b) illustrates the variation detail of 8 SKBs at different external filed, where each bubble displays a random reversal behavior. However, the statistical chirality of the SKBs discloses a collective effect if the in-plane magnetic field varies (Fig. 3(c)). When the negative field was applied and reduced to zero, more C = -1 SKBs were formed. Once reversing the field direction, C = +1 SKBs prefer to exist. Further investigations of another 86 SKBs with more switch processes were shown in Figs. 3(d-f). These analyses qualitatively confirmed that the statistical chirality could not be reversed if the external in-plane field only varied in one direction whatever the strength was large or small.

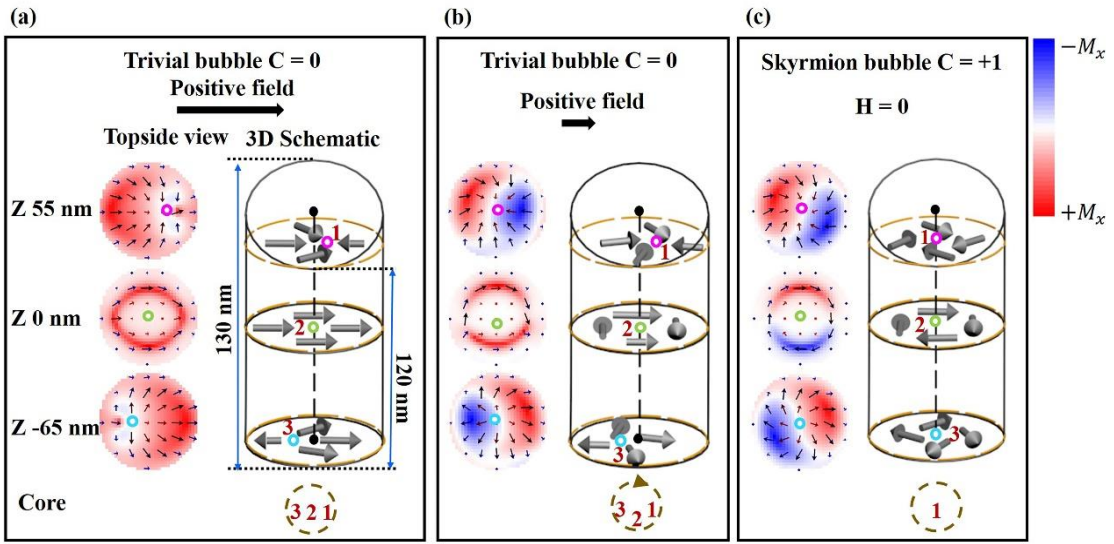

**FIG. 4.** (a-c) Micromagnetic structures of magnetization at different in-plane magnetic fields on a wedge-shaped plate, where the thinner edge faces to the readers. The domain state is obtained from micromagnetic simulation. (a) The left column shows the topside views of the in-plane magnetization of the skyrmion bubble with arrow-pointing orientation and colorized $M_x$ strength at three depth positions (Z = 55 nm, 0 nm, -65 nm). The locations of the magnetization cores at different depth are denoted by (purple-1, green-2 and blue-3 circles in the 3D schematic bubble). The dot line is the cylinder axis. The projection of the cores in different x-y plane is drawn below the cylinder. (b), (c) The redistribution of the magnetization is detailed at the transformation process after removing the in-plane magnetic field.

The modeling and simulation for the chirality reversal of the SKB after applied with in-plane magnetic field was discussed. Considering that the chirality preference mainly occurred by fading the in-plane magnetic field, we performed the OOMMF to systematically explore the mechanism behind the statistic chiral reversal of the SKB in a wedge-shape slab after removing the in-plane magnetic field (Fig. (4)). The achiral bubble (C = 0) emerges when the SKB is stimulated by a higher in-plane magnetic field, as shown in Fig. 4(a). The magnetic spins reveal the twisting Néel domains with the counter swirling at the upper and bottom surfaces of the bubble. Inside of the cylindrical feature, the magnetic spins rearrange themselves to form two Bloch arches. Such variation of the magnetic spins results in the location distortion of the cores along the axis, as demonstrated in the projection image in Fig. 4(a). As the in-plane magnetic field decreases, the magnetic spins at the top and bottom surface maintain the twisting features, but some inner magnetic spins begin to change their directions (Fig. (4b)). The spins in the thinner part attenuate their $M_x$ component but the magnetization in the thicker part keeps the initial direction due to the different static magnetic energy related with the thickness, which has been demonstrated by the further experimental analysis (see S5 within the Supplemental Material [31]). Meanwhile, the cores exhibit the clockwise rotation at the weakened magnetic field, analogous to the SKB chirality. When the in-plane magnetic field disappears, the spins in the thinner part reverse completely and just lead to the energetically favorable C = +1 SKB, while the magnetic spins at the top and bottom still remain the Néel twisting. At the same time, the cores return to the cylinder axis (Fig. 4(c)).

Some characters during the single SKB chirality reversal can be summarized below. First, the magnetization cores at different position of the bubble axis form a swirl, which keeps pace with the SKB chirality during chirality recover. Second, the chirality transition mainly occurs inside the bubble despite the Néel twisting at both surfaces. This behavior can be explained from the perspective of the energy. The total energy of the given system contains the exchange energy, the anisotropy energy, the Zeeman energy, and the demagnetization energy, which are shown in Fig. S6 (see S6 within the Supplemental Material [31]). As the in-plane magnetic field varies, the demagnetizing field prefers to align the magnetic spins in plane whereas the exchange interaction favors the magnetic spins aligned to each other. As a result, the magnetic spins at the surfaces are extremely hard to reverse; as for the interior, the magnetic spins are mainly affected by the Zeeman field, which forces the magnetization along the external field direction. However, these Néel twisting cannot be observed in the LTEM images because the image contrast is dominated chiefly by the inner magnetization owning to the projection characteristic of imaging process (see S7 within the Supplemental Material [31]).Third, the chirality preference arises from the thickness variation in the wedge-like sample. The spins in the thinner part change their orientation easily compared with the thicker part when the external field decreases, which further determines the chirality of the SKB.

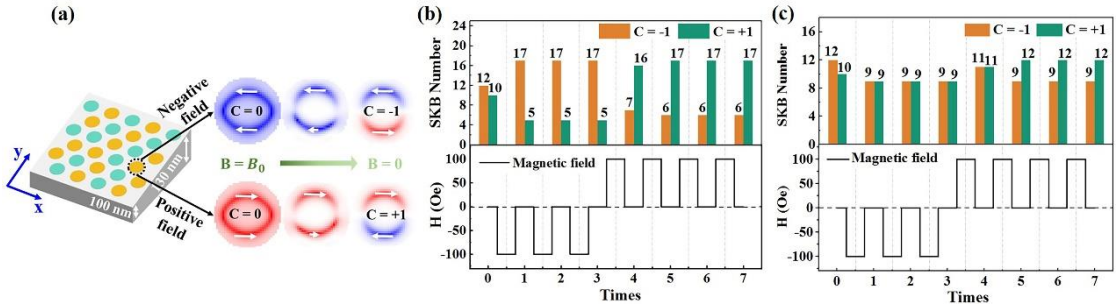

**FIG. 5.** Analyses of chirality reversal calculation with external in-plane magnetic field. (a) Sketch of the field-dependent evolution for skyrmion bubbles (C = ±1) and achiral bubbles (C = 0). $B_0$ is a critical magnetic field where a single skyrmion bubble transform into achiral bubble. The magnetic field is applied along the –x and –y direction. The white arrows represent the in-plane magnetization. The surface size of wedge-like slab is 1000 nm × 1000 nm and the thickness changes between 100 and 130 nm. (b) The reversal statistic for in-plane magnetic filed perpendicular to the thickness gradient and (c) for the magnetic filed along the gradient direction.

As was observed in the experiment, here, we have further performed a micromagnetic simulation of the SKBL with a wedge-shaped slab under a combination of in-plane magnetic field. The simulation manifests some complicities. The schematic field-dependent evolution of SKBs (C = ±1) and achiral bubbles (C = 0) in a wedge-shape slab is shown in Fig. 5(a), where $B_0$ is the critical field to degenerate the chiral or the achiral states of a single bubble. First, a large negative in-plane field (along -x direction) perpendicular to the thickness gradient transforms the SKBs into the achiral bubbles. As the field decreases, the achiral bubbles change into the C = -1 SKBs following the above described transition. If B fades from the positive direction, the procedure is inversed and the C = +1 SKBs are more favorable. Fig. 5(b) displays the statistical distribution of the simulated SKBL, which is consistent with the experimental results in Figs. 3(d-f). In principle, the chirality of the SKBs can be totally controlled by changing the direction of the in-plane magnetic field. However, the magnetic dipolar interaction between SKBs cannot be neglected, which influences the chirality of some SKBs further modifying the final statistical results. If the magnetic field is applied parallel to the thickness gradient direction, or along y direction, different features appear: the chirality reversal randomly and no special chirality prefers (Fig. 5(c)). For comparison, same operation was simulated in another flat lab (Fig. S8) (see S8 within the Supplemental Material [31]). Unlike the wedge-shaped thin plate, the C = ±1 SKBs maintain almost same quantity during the in-plane magnetic field modulation. These simulations demonstrate that the thickness variation in the materials should be an additional freedom to manipulate the chirality of the SKBs if combined with the in-plane magnetic field coding, providing a new route to the device applications for magnetic skyrmions.

**IV. SUMMARY**

In conclusion, the spin chirality reversal of SKBs have been directly observed in the centrosymmetric magnets MnNiGa by using LTEM at room temperature. Remarkably, the in-plane magnetic field can activate the quantity of SKBs reversal. Micromagntic simulations demonstrate that the collective behavior of spin chirality can be strongly affected by the sample thickness when the direction of the magnetic field is inversed. The static energy difference between the thinner and thicker parts within a bubble, associating with the coupling among the bubbles, results in the reversal behavior of the chirality. These findings provide a basic mechanism for manipulating the spin chirality of SKBs in the centrosymmetric magnets, and may also lead to a promising application in nonvolatile spintronic devices with reduced power dissipation.

To date, the key operation of skyrmion racetrack memory is to move skyrmion to read/write by the electrical currents [36]. While promising for technological application, current-driven skyrmion motion is intrinsically collective and only feasible in conductive materials accompanied by undesired heating effects [37]. Here, we experimentally demonstrate that a non-contacting in plane magnetic field can be used to manipulate zero field skyrmion bubble efficiently. In this way, the skyrmion material is no longer detrimentally perturbed by the source of manipulation. The required energy is also significant reduced compared to the current driven-scheme, and it can be applied to all skyrmion-hosting materials regardless of their conductivity. Such applications can also control of their local creation and annihilation through the existence of a skyrmion bubble is thereby interpreted as a '1' and '-1' and its absence as a '0' bit, which can be further induced by the in-plane magnetic field.

Moreover, as devices are surface-dominated, the influence of surfaces on the spin structure is a key question. While a wide range of methods exist for studying 2D skyrmion bubbles, the comprehensive knowledge of 3D magnetic dynamics has remained elusive. Here, combined with micromagntic simulations, we reveal a non-uniform and nontrivial dynamic transition on the surfaces and center of the SKBs and achiral bubbles. Importantly, we reported asymmetric chirality manipulation which can be further explained from the perspective of static energy. Nevertheless, the static energy affects the SKB chirality cannot to be accurately described by the theory at present. We hope our work might evoke more experimental and theoretical studies in this field.

**ACKNOWLEDGEMENTS**

This work was supported by the National Key R&D Program of China (Grant No. 2017YFA0303202, 2017YFA206303), National Natural Science Foundation of China (Grant Nos. 11604148, 11874410), and the Key Research Program of the Chinese Academy of Sciences, KJZD-SW-M01.